\documentclass[12pt,aps,showpacs,eqsecnum,nofootinbib,floatfix]{revtex4}
\usepackage{bbding}
\usepackage{pifont}
\usepackage{subfigure}
\usepackage{epsfig}
\usepackage{CJK}
\usepackage{amsmath}
\usepackage{amsfonts}
\usepackage{amssymb}
\usepackage{color}
\usepackage{graphicx}
\usepackage{mathrsfs}

\def\be{\begin{equation}}
\def\ee{\end{equation}}
\def\ba{\begin{eqnarray}}
\def\ea{\end{eqnarray}}
\def\no{\nonumber}

\def\d{\frac}

\definecolor{dyellow}{rgb}{1.,0.8,.0}
\definecolor{myblue}{rgb}{.1,.1,.7}
\definecolor{dcyan}{rgb}{.0,.6,.6}
\definecolor{dmagenta}{rgb}{0.6,0.0,0.6}
\definecolor{brown}{rgb}{0.6,0.2,0.}
\definecolor{darkblue}{rgb}{.0,.0,0.5}
\definecolor{darkred}{rgb}{0.75,0.0,0.0}
\definecolor{orange}{rgb}{1.,.6,.0}
\definecolor{dorange}{rgb}{0.8,.4,.0}
\definecolor{darkgreen}{rgb}{0.0,0.6,0.0}
\definecolor{purple}{rgb}{.4,.0,.4}
\definecolor{lightgrey}{rgb}{0.7, 0.7, 0.7}
\definecolor{grey}{rgb}{0.4, 0.4, 0.4}

\def\d#1#2{\frac{\displaystyle #1}{\displaystyle #2}}

\newcommand\btd{\raise 2pt
\hbox{$\hat\bigtriangledown$}\hskip 1.5pt}
\newcommand\bt{\raise 2pt
\hbox{$\bigtriangledown$}\hskip 1.5pt}

\newcommand{\omits}[1]{}

\def\PRD{{Phys. Rev.}~{\bf D}}

\def\CQG{{Class. Quant. Grav. }}

\def\JHEP{{JHEP}}

\def\IJMPA{{Int. J. Mod. Phys.}~{\bf A}}

\begin{document}

\title{Critical behaviors of black hole in an asymptotically safe gravity with cosmological constant}

\author{Meng-Sen Ma$^{a,b}$\footnote{Email: mengsenma@gmail.com}, Ya-Qin Ma$^{c}$}

\medskip

\affiliation{\footnotesize$^a$Department of Physics, Shanxi Datong
University,  Datong 037009, China\\
\footnotesize$^b$Institute of Theoretical Physics, Shanxi Datong
University, Datong 037009, China\\
\footnotesize$^c$Medical College, Shanxi Datong
University,  Datong 037009, China}

\begin{abstract}

We study the $P-V/r_{+}$ criticality and phase transition of quantum-corrected
black hole in asymptotic safety (AS) gravity in the extended phase space. For the black hole, the cosmological constant is dependent on the
momentum cutoff or energy scale, therefore one can naturally treat it as a variable and connect it with the thermodynamic pressure.
We find that for the quantum-corrected black hole there is a similar first-order phase transition to that of the van der Waals liquid/gas system.
We also analyze the types of the phase transition at the critical points according to Ehrenfest's classification.
It is shown that they are second-order phase transition.
\end{abstract}

\maketitle

\section{introduction}

Like ordinary thermodynamic matter, black holes also have temperature, entropy and energy.
The laws of black hole mechanics have the similar forms to the laws of thermodynamics\cite{Hawking1}. Therefore,
we can treat black holes as thermodynamic systems. In fact, between black holes and the conventional thermodynamic systems,
 there are other similarities, such as phase transition and critical behaviors. The pioneering work of Davies\cite{Davies} and
 the well-known Hawking-Page phase transition\cite{Hawking2} are both proposed to elaborate these points. The phase transitions and critical
  phenomena in anti-de Sitter (AdS) black holes have been studied extensively\cite{Hut,Mazur,Lousto,Lemos,Banerjee1,Banerjee2,Myung}. Some interesting works
show that there exists phase transition similar to the van der Waals
liquid/gas phase transition for some black holes\cite{chamblin,chamblin1,Wu,Kastor,Cvetic,
Tian,Banerjee3,LYX,BM}.  Even for the black holes in
  dS space critical behaviors can also be studied by considering the connections between the black hole horizon and
  the cosmological horizon\cite{Zhao1,Zhao2}.

Recently, some physicists reconsidered the critical phenomena of AdS black holes by treating the cosmological constant $\Lambda$ as
a variable and connecting it with the thermodynamic pressure\cite{Dolan1,RBM,RBM2,RBM3,Cai,Hendi}. In
some models, the cosmological constant may be considered to be a
time-variable quantity\cite{Peebles,RH}, or as some thermodynamic
quantities, such as thermodynamic pressure\cite{Brown1,Brown2},
which should be a conjugate quantity of thermodynamic volume.
Inclusion of the variation of $\Lambda$ can make the first law of black hole thermodynamics   consistent with the Smarr formula for some black holes.

In this letter, we study the critical behaviors of a kind  black hole derived in asymptotic safety gravity. The asymptotic safety scenario for quantum gravity was
put forward by Weinberg\cite{Weinberg}. It is based on a nontrivial
fixed point of the underlying renormalization group(RG) flow for gravity. This theory has been studied extensively and applied to several
 different subjects in quantum gravity\cite{Reuter1,Niedermaier,Reuter2}. Bonanno and Reuter\cite{Reuter3} derived
 the renormalization group improved black hole metrics by replacing Newtonian coupling constant with  a `` running " one. Cai, et.al \cite{CYF} find
a spherically symmetric vacuum solution to field equation derived from the AS gravity with higher derivative terms and with cosmological constant.
In this theory, the cosmological constant is no longer constant but dependent on a momentum cutoff. Therefore it is reasonable to include the variation of $\Lambda$ in the first law of black hole thermodynamics as thermodynamic pressure $P$. The quantum correction of AS gravity to the conventional Schwarzschild-AdS black hole makes its thermodynamic quantities and critical behaviors very different. The quantum-corrected black hole can also show a phase transition analogous
to the liquid-gas phase transition in the Van der Waals system.
According to Ehrenfest's classification we also consider the Gibbs free energy, the isothermal
 compressibility and the expansion coefficient. It is shown that the type of phase transition for the  black hole at the critical point belongs to  the second-order or continuous one.

The paper is arranged as follows: in the next section we simply
introduce the AS gravity model and its quantum-corrected black hole solution. In section 3 we will study the $P-V/r_{+}$ criticality by
considering the cosmological constant as thermodynamic pressure. We also calculate the critical exponents here.  In
section 4  we also analyze the type of the phase
transition of the quantum-corrected black hole in the extended phase space according to
Ehrenfest's classification.
 We  make some
concluding remarks in section 5.

\section{Quantum-corrected black hole in AS gravity}

We start with a generally covariant effective gravitational action with higher derivative terms involving a momentum cutoff $p$\cite{CYF} :
\ba
\Gamma_p[g_{\mu\nu}]&=&\int dx^4\sqrt{-g}~[p^4g_0(p)+p^2g_1(p)R +g_{2a}(p)R^2+g_{2b}(p)R_{\mu\nu}R^{\mu\nu} \no \\
&+& g_{2c}(p)R_{\mu\nu\sigma\rho}R^{\mu\nu\sigma\rho}+ O(p^{-2}R^3)+ ...],
\ea
where $g$ is the determinant of the metric tensor $g_{\mu\nu}$, $R$ is the Ricci scalar, $R_{\mu\nu}$ is the Ricci tensor and $R_{\mu\nu\sigma\rho}$ is
the Riemann tensor. The coefficients $g_{i} (i=0,1,2a,...)$ are dimensionless coupling parameters and are functions of the dimensionful, UV cutoff.
The couplings satisfy the RG equations:
\be
\d{d}{d \ln p}g_{i}(p)=\beta_{i}[g(p)]
\ee
Assuming a static spherically symmetric metric ansatz and choosing the Schwarzschild gauge
\be
ds^{2}=-f(r)dt^{2}+f(r)^{-1}dr^{2}+r^{2}d\Omega_2 ^{2},
\ee
and then substituting it into the generalized Einstein field equations
\be
\tilde{G}^{\mu\nu}\equiv \d{\delta\Gamma_p[g_{\mu\nu}]}{\delta g_{\mu\nu}}=0,
\ee
one can derive a Schwarzschild-(anti)-de Sitter-like solution
\be\label{metric}
f(r)=1-\d{2G_p M}{r}\pm \d{r^2}{l_p^2},
\ee
where $G_p$ and $l_p$ are the gravitational coupling and the radius of the asymptotically (A)dS space, and both depend on the momentum cutoff $p$.

It is shown in \cite{CYF} that there are a Gaussian fixed point in the IR limit and a non-Gaussian fixed point in the UV limit. A central result is
\ba
&&g_0 \simeq -\d{(\Lambda_{IR}+\eta p^2G_{N})(1+\xi p^2G_{N})}{8\pi p^4G_{N}} \\
&&g_1 \simeq \d{1+\xi p^2G_{N}}{16\pi p^2 G_{N}},
\ea
where $G_{N}$ and $\Lambda_{IR}$ are the values of the gravitational coupling and the cosmological constant in the IR limit which should be determined by
astronomical observations. The parameters $\xi$ and $\eta$ are both related to the running couplings $\lambda(p)g_{2a},~\lambda(p)g_{2b},~\lambda(p)g_{2c}$
at the non-Gaussian fixed point. The coefficient $\lambda$ has the familiar logarithmic form which approaches asymptotic freedom
\be
\lambda(p) =\d{\lambda_0}{1+\d{133}{160\pi^2}\lambda_0\ln p/M_p},
\ee
where $\lambda_0$ is a fixed value of the coefficient $\lambda$ at the Planck scale, and $M_p$ is the Planck mass.

It is shown that the running gravitational coupling $G_p$ is related the Newtonian gravitational coupling constant $G_N$ by
\be
G_p=\d{G_N}{1+\xi p^2G_N}.
\ee
At high energy scale,
\be
p(r)\simeq 2.663(\d{M^2}{|\lambda_0|})^{1/8}r^{-3/4}.
\ee
thus,
\be f(r)\simeq 1-\d{625}{512\pi}|\lambda_0|^{1/4}(Mr)^{1/2}
\ee
which is singular-free at $r=0$. However, the curvature singularity still exists due to  divergent
 $R_{\mu\nu}R^{\mu\nu},~R_{\mu\nu\rho\sigma}R^{\mu\nu\rho\sigma}$.

At low energy scale, the momentum cutoff drop to the infrared(IR) limit, and $p\simeq \d{1}{r}$.
At this time,
\be\label{metric1}
f(r) \simeq 1-\d{2Mr}{r^2+\xi}\pm\d{r^2}{l_{p}^2}
\ee
where $G_N=1$ has been set for simplicity. The parameter $\xi$ represents the quantum correction to the conventional Schwarzschild-AdS
black hole. Obviously, when $\xi=0$, the corrected metric will return back to the Schwarzschild-AdS one.
Below we will study the thermodynamics of the quantum-corrected black holes based on Eq.(\ref{metric1}). It is shown that owing to the correction,
 the thermodynamic quantities will also be corrected.

\section{$P-V$ criticality of the quantum-corrected black hole}

In this paper we only concern with the asymptotic AdS black hole. Firstly, we identify the pressure with
\be
 P=\d{3}{8\pi l_{p}^2}.
 \ee
 From Eq.(\ref{metric1}), one can easily obtain the mass
 \be
 M=\frac{\left(8 \pi  P r_{+}^2+3\right) \left(\xi +r_{+}^2\right)}{6 r_{+}},
 \ee
 where $r_{+}$ is the radius of the black hole event horizon.

The first law of black hole thermodynamics should written as
\be\label{1st}
dM=TdS+VdP
\ee
where the conjugate thermodynamic volume $V=\left.\d{\partial M}{\partial P}\right|_{S}=\frac{4}{3} \pi  r_{+} \left(\xi +r_{+}^2\right)$.
Here, the mass of black hole is no more internal energy, but should be interpreted as the thermodynamic
enthalpy, namely $H=M(S,P)$\cite{Kastor,Dolan1,Dolan2,Dolan3}. The first law of black hole thermodynamics represented by the internal energy $U(S,V)$ reads
\be\label{flenergy}
dU=TdS-PdV
\ee
where $U=H-PV$.

The Hawking temperature of the black hole can be easily derived
\be\label{temp}
T=\d{f'(r_{+})}{4\pi}= \frac{-3 \xi +24 \pi  P r_{+}^4+8 \pi  \xi  P r_{+}^2+3 r_{+}^2}{12 \pi  r_{+}^3+12 \pi  \xi  r_{+}}.
\ee
When $\xi=0$, it will give the temperature of Schwarzschild-AdS black hole. According to the first law, Eq.(\ref{1st}), one can derive the entropy
\ba\label{entropy}
S=\int \d{d M}{T}=\int \left.\d{1}{T}\d{\partial M}{\partial r_{+}}\right|_{P} dr_{+}= \pi r_{+}^2+2\pi \xi\ln \d{r_{+}}{\sqrt{\xi}}+S_0
\ea
where $S_0$ is an integration constant which can be decided by the boundary conditions. The additional logarithmic term in
the expression of the entropy indicates the quantum gravitational correction. As the parameter $\xi\rightarrow 0$, the standard Bekenstein-Hawking
area law will return. It is interesting that no $P$ exists in the expression of $S$ although it is included in $M$ and $T$.
\begin{figure}[!htbp]
\center{\subfigure[] {
\includegraphics[angle=0,width=8cm,keepaspectratio]{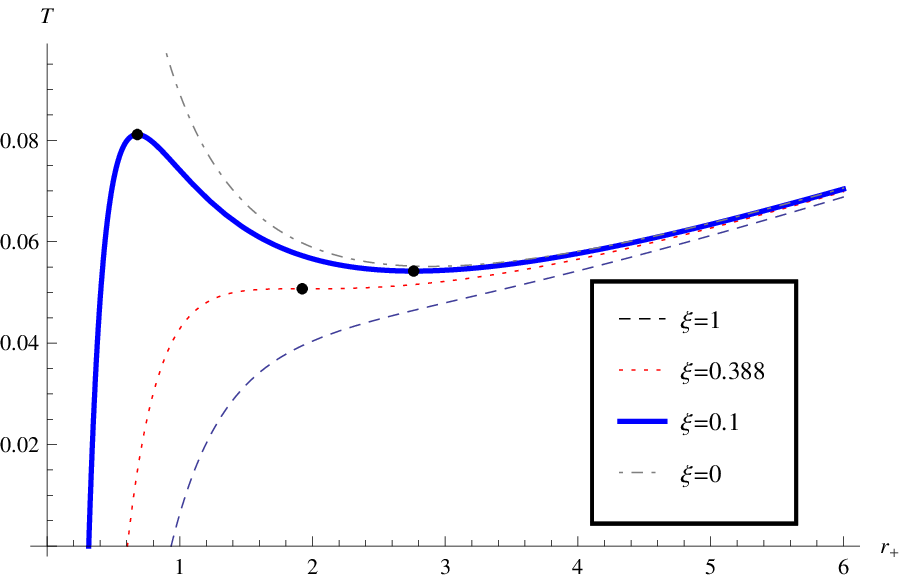}}
\subfigure[] {
\includegraphics[angle=0,width=8cm,keepaspectratio]{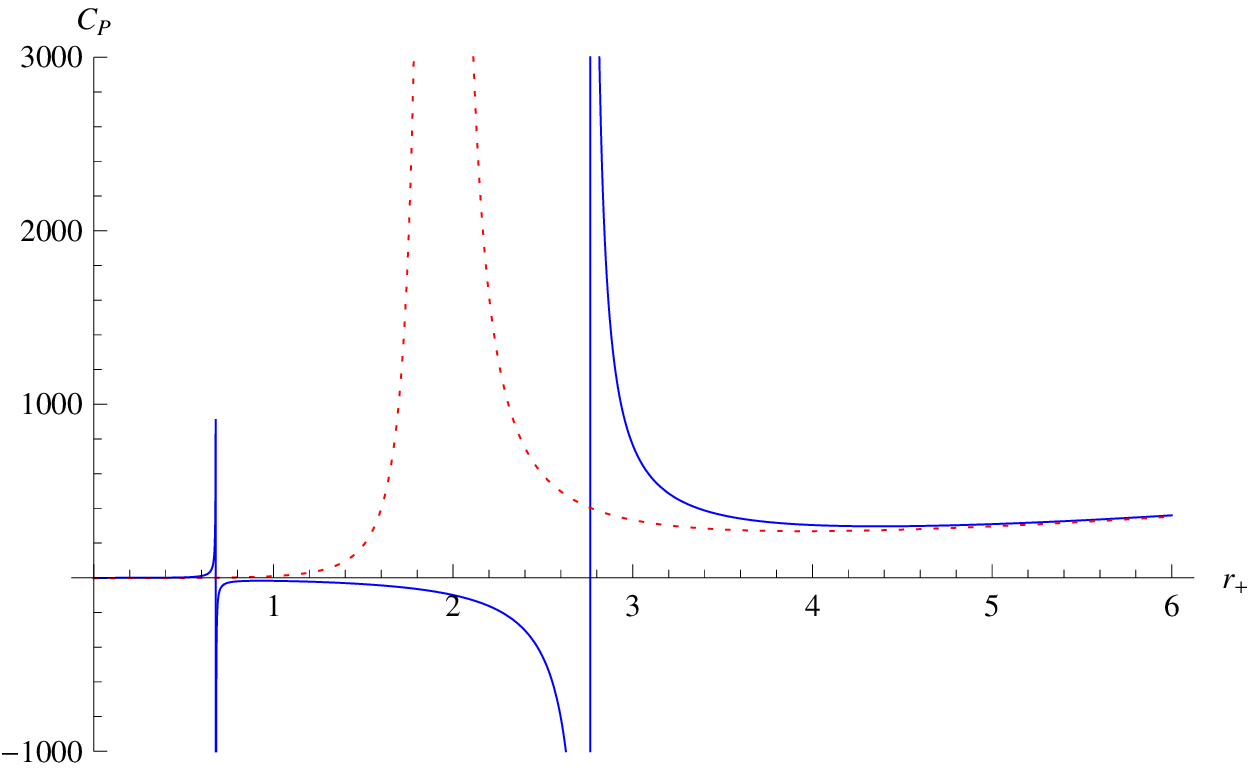}}
\caption[]{\it The temperature and heat capacity at constant pressure as functions of $r_{+}$. We set $l_{p}=5$ here.
The three black dots represent the local extrema of the temperature. The dot-dashed line in (b) corresponds to the temperature of
the Schwarzschild-AdS black hole.}}\label{TC}
\end{figure}

The heat capacity at constant pressure can be given by
\be
C_{P}=\left.\d{\partial M}{\partial T}\right|_{P}=\frac{2 \pi  \left(\xi +r_{+}^2\right)^2 \left(-3 \xi +24 \pi  P r_{+}^4+8 \pi  \xi  P r_{+}^2+3 r_{+}^2\right)}
{3 \xi ^2+24 \pi  P r_{+}^6+64 \pi  \xi  P r_{+}^4+8 \pi  \xi ^2 P r_{+}^2-3 r_{+}^4+12 \xi  r_{+}^2}
\ee
The qualitative behaviors of the temperature $T$ and the heat capacity $C_{P}$ are depicted in Fig.1.
Obviously, owing to the existence of $\xi$, the temperature will not blow up as the radius of the event horizon approaches zero, but tends to zero at a finite radius where the black hole will become extremal one. For a fixed pressure there is a critical value of $\xi$, below which there will be both local maximum and minimum for the temperature, and above which no local extremum exists. At the critical value, the maximum and minimum will coincide. From Fig.1(b), one can see that, when $\xi<\xi_c$, $C_P$ suffers discontinuities at two points , which can be identified as the critical points for phase transition in the quantum-corrected black hole. The divergences of the heat capacity appear precisely at the extrema of the temperature. The small and large black holes with positive heat capacity can be stable. While the intermediate black hole with negative heat capacity is instable.

From Eq.(\ref{temp}), one can derive the equation of state of the black hole
\be\label{eos}
P=\frac{3 \left(\xi +4 \pi  r_{+}^3 T-r_{+}^2+4 \pi  \xi  r_{+} T\right)}{8 \pi  r_{+}^2 \left(\xi +3 r_{+}^2\right)}
\ee
One can take the specific volume as $v \propto V/N$, with  $N=A/l_P^2$ counting the number of degrees of
freedom associated with the black hole horizon\cite{RBM3}. $l_P$ here is the Planck length. If we take $v=6V/N$, 
the specific volume can be expressed as
\be\label{sv}
v=2(r_{+}+\d{\xi}{r_{+}}).
\ee
Obviously, when $\xi=0$, it will give the result similar to that in \cite{RBM,RBM2}.
Replacing $r_{+}$ in Eq.(\ref{eos}) with $v$, one can obtain the equation of motion, $P=P(v,T,\xi)$.
As is done in \cite{RBM2}, one can also expand $P(v,T,\xi)$ in powers of $a$ in the small $a$ limit and take the first several terms approximately.
\be
P=\d{T}{v}-\d{1}{2\pi v^2}+\d{4(5\pi Tv-1)\xi}{3\pi v^4}+\d{8(68\pi Tv-1)\xi^2}{9\pi v^6}+O[\xi]^3
\ee
One can see that the above equation of state is similar to that of Kerr-AdS black hole. It indeed exhibits $P-v$ criticality.
However, in this paper we want to treat  Eq.(\ref{eos}) exactly. We will use the horizon radius in the equation of the state instead of the specific volume hereafter.

\begin{figure}[!htbp]
\includegraphics[angle=0,width=10cm,keepaspectratio]{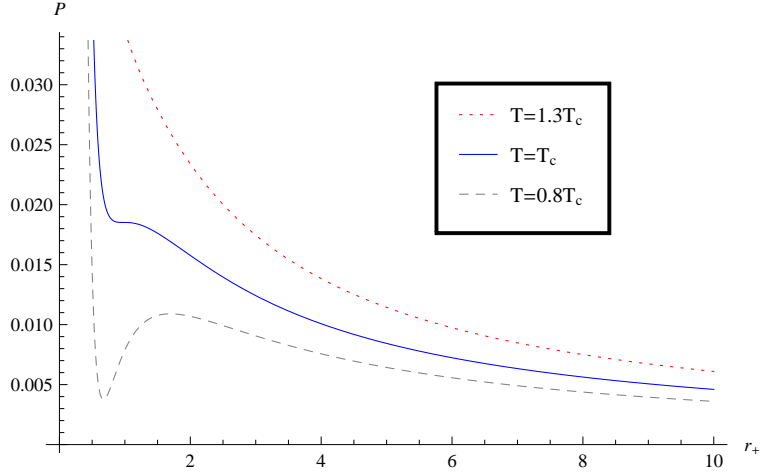}
\caption[]{\it $P-r_{+}$ diagram for the AS improved black hole. We choose $\xi=0.1$ here.
The critical radius, temperature and pressure are respectively  $r_c=0.976, ~ T_c=0.0999,~P_c=0.0185$ .}\label{PV}
\end{figure}

The critical point can be obtained according to
\be
\d{\partial P}{\partial r_{+}}=0, \quad \d{\partial^2P}{\partial r_{+}^2}=0
\ee
which lead to
\be
r_c=\sqrt{c \xi}, \quad T_c=\frac{3 c^2-6 c-1}{2 \pi  \left(3 c^2+8 c+1\right) \sqrt{c \xi }}, \quad P_c=\frac{3 \left(c^2-4 c-1\right)}{8 \pi  c \left(3 c^2+8 c+1\right) \xi }
\ee
where the constant $c=3+\d{2}{3^{2/3}}\left[(39+i\sqrt{15})^{1/3}+(39-i\sqrt{15})^{1/3}\right]$, which is a real number. Numerically $c\approx 9.53$.
These critical values can lead to the following universal ratio
\be
\rho_c=\d{P_cr_c}{T_c}=\frac{3 \left(c^2-4 c-1\right)}{4 \left(3 c^2-6 c-1\right)}\approx 0.181.
\ee
Obviously, it is independent of the quantum-corrected constant $\xi$. Note that for the van der Waals gas, the universal ratio is $\rho_c=3/8$, while
for some actual gas, such as water, it is $\rho_c=0.230$.

Furthermore, one can analyze the Gibbs free energy: $G=G(T,P)=H-TS=M-TS$.
As is shown in Fig.3, the Gibbs free energy develops a `` swallow tail" for $P<P_c$, which is a typical feature in a first-order phase transition.
Above the critical pressure $P_c$, the `` swallow tail" disappears.

\begin{figure}[htbp]
\includegraphics[angle=0,width=8cm,keepaspectratio]{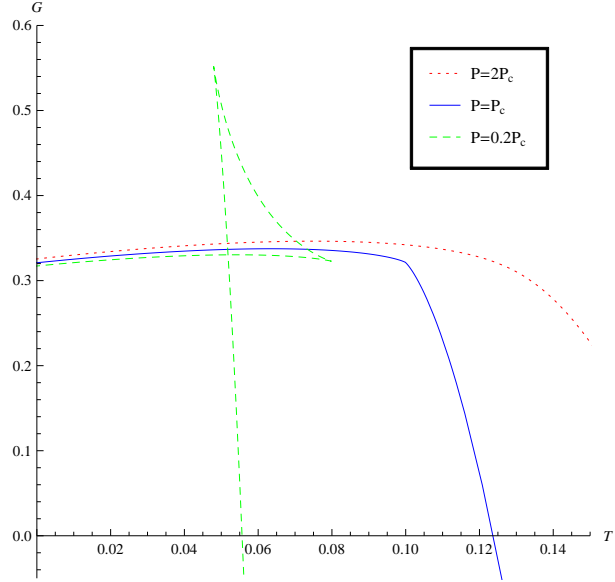}
\caption[]{\it The Gibbs free energy as function of temperature for different pressure for the quantum-corrected black hole. We also choose $\xi=0.1$ here.
}\label{GT}
\end{figure}

Next we will calculate the critical exponents at the critical point for the quantum corrected black hole. For a van der Waals liquid/gas system, the critical behaviors
can be characterized by the critical exponents as follows\cite{Stanley}:
\be
C_v\sim \left(-\d{T-T_c}{T_c}\right)^{-\alpha}, \quad \d{v_g-v_l}{v_c} \sim \left(-\d{T-T_c}{T_c}\right)^{\beta},
\quad \kappa_{T}\sim \left(-\d{T-T_c}{T_c}\right)^{-\gamma}, \quad P-P_c \sim (v-v_c)^{\delta}.
\ee
Here $v_g$ and $v_l$ refer to the specific volume for gas phase and liquid phase respectively. For the quantum-corrected black hole,
we  use $r_g$ and $r_l$ instead. Defing
\be
t=\d{T}{T_c}-1, \quad x=\d{r_{+}}{r_c}-1, \quad p=\d{P}{P_c}
\ee
and replacing $r_{+},~T,~P$ in Eq.(\ref{eos}) with the new dimensionless parameters $x,~t,~p$ and then expanding the equation near the critical point approximately, one can obtain
\be\label{app}
p=1+At+Btx+Cx^3+O(tx^2,x^4),
\ee
where $A,~B,~C$ are all complicated expressions composed of the $c=3+\d{2}{3^{2/3}}\left[(39+i\sqrt{15})^{1/3}+(39-i\sqrt{15})^{1/3}\right]$. Numerically, $A\approx 2.95, ~ B\approx -3.31, ~ C \approx -1.20$. Eq.(\ref{app}) has the same form as that for the van der Waals system and the RN-AdS black hole\cite{RBM}. Therefore, we can derive
the critical exponents $\beta=1/2,~\gamma=1,~ \delta=3$ in the same way. In addition, according to Eq.(\ref{entropy}), the entropy is independent of $T$. Thus,
$C_V=T\left.\d{\partial S}{\partial T}\right|_V=0$. Therefore, we also have the critical exponent $\alpha=0$.
Obviously, they obey the scaling symmetry like the ordinary thermodynamic systems:
\ba
&&\alpha+2\beta+\gamma=2, \quad \alpha+\beta(\delta+1)=2 \no \\
&&\gamma(1+\delta)=(2-\alpha)(\delta-1), \quad \gamma=\beta(\delta-1).
\ea

\section{The second-order  phase transition at the critical point}

In this section, we study the types of the phase transition for the quantum-corrected black hole at the critical points. It should be noted that
the critical points depend on the values of the pressure $P$ or the temperature $T$ for a positive $\xi$. When $P=P_c$ or $T=T_c$, there is only one critical point;
when $P<P_c$ or $T<T_c$ there will be two critical points; no critical point exists when $P>P_c$ or $T>T_c$.

Ehrenfest had ever attempted to classify the phase transitions. Phase transitions connected with
  an entropy discontinuity are called discontinuous or first order phase transitions, and phase transitions where the entropy is continuous are called continuous or second/higher order phase transitions. More precisely, for the first-order phase transition the Gibbs free energy $G(T,P,...)$ should be continuous and
  its first derivative with respect to the external fields:
  \be
  S=-\left.{\frac{\partial G}{\partial T}}\right|_{(P,...)}, \quad  V=\left.{\frac{\partial G}{\partial P}}\right|_{(T,...)}
  \ee
  are discontinuous at the phase transition points.

 For the second-order phase transition the Gibbs free energy $G(T,P,...)$ and its first derivative are both continuous,
 but the second derivative of $G$ will diverge at the phase transition points, such as the specific heat $C_P$,
 the compressibility $\kappa$, the expansion coefficient $\alpha_{P}$:
 \be
 C_P=T\left.{\frac{\partial S}{\partial T}}\right|_P=-T\left.{\frac{\partial^2 G}{\partial T^2}}\right|_P,
 \kappa_{T} =-\frac{1}{V}\left.{\frac{\partial V}{\partial P}}\right|_T=-\frac{1}{V}\left.{\frac{\partial^2 G}{\partial P^2}}\right|_T ,
 \alpha_{P} =-\frac{1}{V}\left. {\frac{\partial V}{\partial T}} \right|_P=-\frac{1}{V}\frac{\partial^2 G}{\partial P\partial T}
 \ee

\begin{figure}[!htbp]
\centering
\includegraphics[angle=0,width=8cm,keepaspectratio]{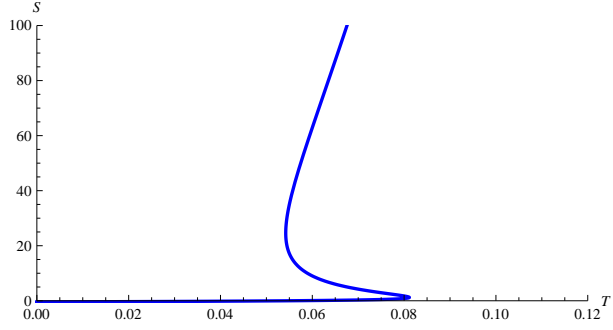}
\caption[]{\it The entropy as functions of temperature for the choices of $ ~\xi=0.1, ~l_{p} =5$.}
\label{ST}
\end{figure}

According to Eq.(\ref{temp}) and Eq.(\ref{entropy}), one can easily obtain the $S-T$ plot, as is shown in Fig.\ref{ST}. Obviously, the entropy is a continuous
function of temperature.

\begin{figure}[!htbp]
\includegraphics[angle=0,width=7cm,keepaspectratio]{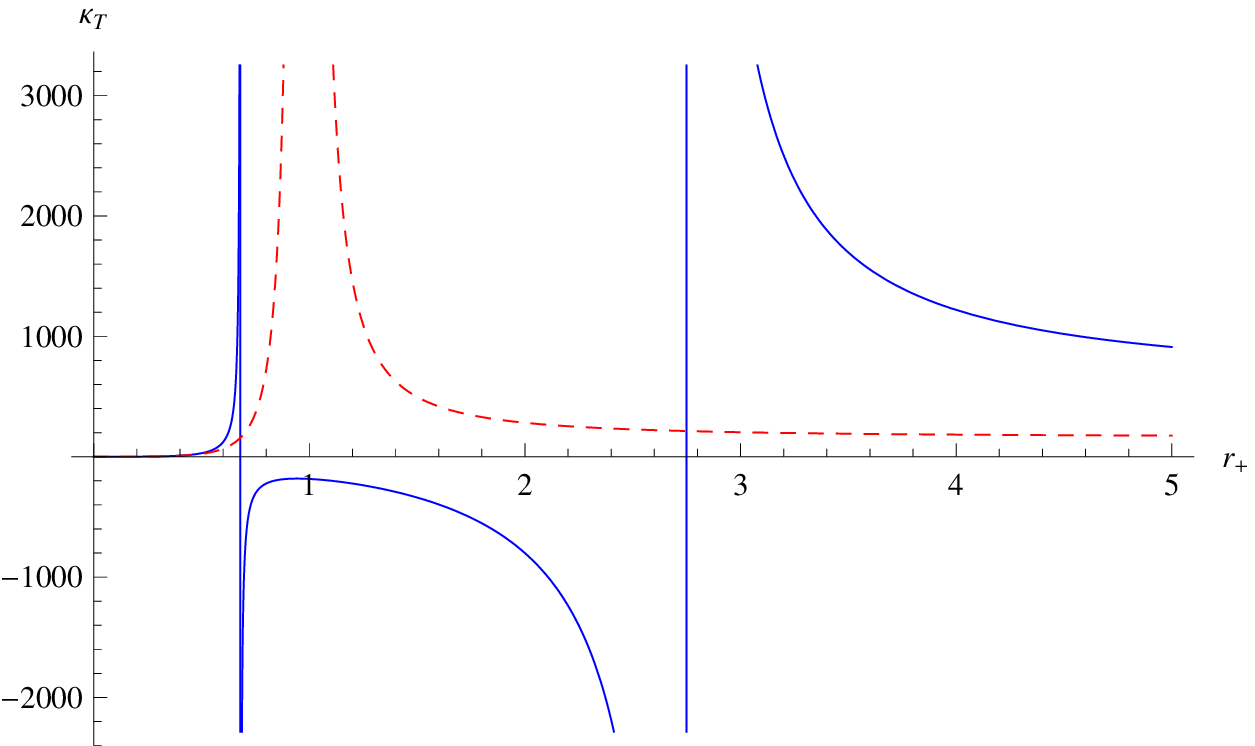}\hfill
\includegraphics[angle=0,width=7cm,keepaspectratio]{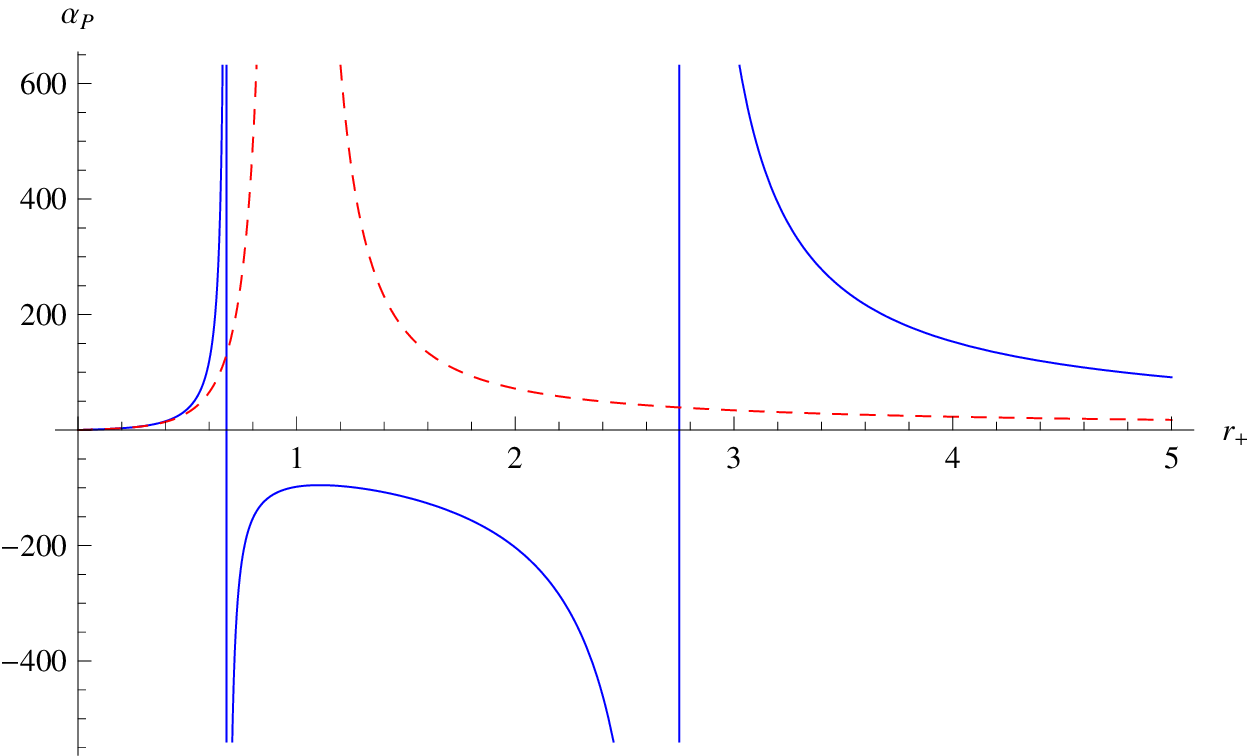}
\caption[]{\it The compressibility $\kappa$ and the expansion coefficient $\alpha_{P}$
 as functions of $r_{+}$ for the quantum-corrected black hole. The solid (blue) curve corresponds to $P=0.26P_c$, and the dashed (red) curve corresponds to
 $P=P_c$. We also set $\xi=0.1$. }
\label{KAT}
\end{figure}

One can easily calculate the $\kappa_{T}$ and $\alpha_{P}$:
\ba
\kappa_{T} &=&-\frac{1}{V}\left.\frac{\partial V}{\partial r_{+}}\frac{\partial r_{+}}{\partial P}\right|_T=
\frac{8 \pi  r_{+}^2 \left(\xi +3 r_{+}^2\right)^2}{3 \xi ^2+24 \pi  P r_{+}^6+64 \pi  \xi  P r_{+}^4+8 \pi  \xi ^2 P r_{+}^2-3 r_{+}^4+12 \xi  r_{+}^2}\\
\alpha_{P} &=&\frac{1}{V}\left.\frac{\partial V}{\partial r_{+}}\frac{\partial r_{+}}{\partial T}\right|_P=
\frac{12 \pi  r_{+} \left(\xi +r_{+}^2\right) \left(\xi +3 r_{+}^2\right)}{3 \xi ^2+24 \pi  P r_{+}^6+64 \pi  \xi  P r_{+}^4+8 \pi  \xi ^2 P r_{+}^2-3 r_{+}^4+12 \xi  r_{+}^2}
\ea
They will diverge when the denominator vanishes. It is clear that the denominators of the $\kappa_{T},~\alpha_{P}$ are the same as that of $C_P$.
As shown in Fig.\ref{KAT}, there will be two divergent points for both $\kappa_{T}$ and $\alpha_{P}$ for $P<P_c$. Only one divergent point left when
$P=P_c$. Owing to the divergence of $\kappa_{T},~\alpha_{P}$, phase transitions at these critical points are all second-order.

When $\xi=0$, the quantum-corrected black hole return to the Schwarzschild-AdS black hole, for which there is still critical point where
$C_{P},~\kappa,~\alpha_{P}$ diverge. However, in this case, only one critical point exists.
One can also analyze the types of the phase transition at the critical
point by means of Ehrenfest scheme employed in \cite{Banerjee4}. That can give the same result.
Generally, thermodynamic geometry can also be employed to study the phase transition\cite{Ruppeiner,Quevedo1,Quevedo2}. However, for
the quantum-corrected black hole it does not work. Because the mass/enthalpy is linear in the pressure $P$, which will lead to
 degenerate thermodynamic metric.

\section{concluding remarks}

In this paper we studied the thermodynamics and critical behaviors of a kind of quantum-corrected black hole obtained
in the asymptotically safe gravity theory with higher derivatives and cosmological constant.
The asymptotic safety scenario includes the scale dependent Newtonian `` constant" $G_p$ and cosmological `` constant".
$G_p$ leads to the correction to the conventional Schwarzschild-AdS black hole. The running cosmological `` constant" can be
treated as a variable naturally. We can identify it with the thermodynamic pressure and include its variation in the first law of black hole thermodynamics.

Based on the quantum-corrected black hole, we studied the $P-V/r_{+}$ criticality at the critical point and plotted the isotherm curves. It is shown that
the $P-V/r_{+}$ phase diagram is the same as that of the van der Waals liquid/gas system. Furthermore, we calculated the critical exponents at the
critical point, which all coincide with that of the van der Waals system and RN-AdS black hole. From the critical parameters we can also construct
 the universal ratio $\rho_c=\d{P_c r_c}{T_c}\approx 0.181$.
We analyzed the types of phase transition at the critical points  using Ehrenfest's classification.
The Gibbs free energy and entropy are both continuous  functions of temperature. The heat capacity at constant pressure $C_{P}$,
the compressibility $\kappa_{T}$ and the expansion coefficient $\alpha_{P}$ all suffer discontinuities at some points when the pressure or the temperature
is not larger than their critical values. Therefore, we conclude that the phase transitions at these points belong to the second-order one.

\bigskip

\section*{Acknowledgements}
MSM thanks Prof. Ren Zhao for useful discussion. This work is supported in part by NSFC under Grant
Nos.(11247261;11175109;11075098;11205097).

\end{document}